\documentclass[aps,prl,numerical,superscriptaddress,showpacs,floatfix,reprint]{revtex4-1}

\usepackage{graphicx,color}
\usepackage{amsmath,amssymb,amsfonts}
\usepackage{hyperref}

\def \be {\begin{equation}}
\def \ee {\end{equation}}
\def \ee  {\end{equation}}
\def \bea {\begin{eqnarray}}
\def \eea {\end{eqnarray}}

\begin{document}
\title{
A model investigation of the longitudinal broadening of the \\transverse momentum two-particle correlator
}

\medskip

\author{Niseem~Magdy} 
\email{niseemm@gmail.com}
\affiliation{Department of Physics, University of Illinois at Chicago, Chicago, Illinois 60607, USA}


\author{Roy~A.~Lacey} 
\affiliation{Department of Chemistry, State University of New York, Stony Brook, New York 11794, USA}


\begin{abstract}
The Multi-Phase Transport model (AMPT) is used to investigate the longitudinal broadening of the 
transverse momentum two-particle correlator $C_{2}\left(\Delta\eta,\Delta\varphi\right)$, and its utility to extract the specific 
shear viscosity, $\eta/s$, of the quark-gluon plasma formed in ultra-relativistic heavy ion collisions.
The results from these model studies indicate that  the longitudinal broadening 
of $C_{2}\left(\Delta\eta,\Delta\varphi\right)$ is sensitive to the value of $\eta/s$. However, 
reliable extraction of the longitudinal broadening of the correlator requires the suppression of 
possible self-correlations associated with the definition of the collision centrality.

\end{abstract}

\maketitle

A primary aim of the heavy-ion programs at the Relativistic Heavy-Ion Collider (RHIC) and the Large Hadron Collider (LHC) is to understand the properties of the quark-gluon plasma (QGP)~\cite{Shuryak:1978ij,Shuryak:1980tp,Muller:2012zq} formed in high-energy heavy ion collisions.  In particular, understanding the QGP transport properties, especially the specific shear viscosity ($\eta/s$), i.e., the ratio of shear viscosity $\eta$, to the entropy density $s$, which characterizes the ability of the QGP to transport momentum, has gained broad consideration.
 
The azimuthal anisotropy of particle emission in the transverse plane, recognized as anisotropic flow, is one of the main observables used to constrain the viscous hydrodynamic response (reflecting $\eta/s$) to the initial spatial distribution in energy density, created in the early stages of the collision~\cite{Danielewicz:1998vz,Ackermann:2000tr,Adcox:2002ms,Heinz:2001xi,Hirano:2005xf,Huovinen:2001cy,Hirano:2002ds,Romatschke:2007mq,Luzum:2011mm,Song:2010mg,Qian:2016fpi,Schenke:2011tv,Teaney:2012ke,Gardim:2012yp,Lacey:2013eia,Magdy:2020gxf}. 
Initial theortical model comparisons to flow measurements at RHIC and LHC suggested a small $\eta/s$ value for the QGP~\cite{Shuryak:2003xe,Romatschke:2007mq,Luzum:2008cw,Bozek:2009dw,Song:2010mg,Shen:2011eg},
albeit with sizable uncertainties stemming from the uncertainty in the initial-state eccentricities.
 
Subsequently, several theoretical and experimental investigations have been devoted to the development of further constraints for more robust extractions of $\eta/s$~\cite{Schenke:2019ruo,Alba:2017hhe,Gonzalez:2020bqm,Everett:2020xug,Gardim:2020mmy,Gonzalez:2020gqg,Agakishiev:2011fs,Acharya:2019oxz,Kovtun:2004de}. Although these investigations have advanced the precision of the extraction of $\eta/s$~\citep{Chatrchyan:2013kba,Sirunyan:2019izh,ALICE:2016kpq,Adam:2020ymj,Niemi:2015qia,Danielewicz:2002pu,Luzum:2010fb,Teaney:2010vd,Adams:2005ca,Magdy:2019ojv,Adamczyk:2016gfs,Ollitrault:2009ie,Adam:2019woz,Magdy:2018itt, Adamczyk:2017ird,Magdy:2017kji,Adamczyk:2017hdl,Alver:2010gr,Magdy:2020bij,Adare:2011tg,Adamczyk:2013waa,Acharya:2017zfg}, more stringent constraints are still required for the initial-state \cite{Song:2010mg,Qiu:2011hf,Song:2012tv} and the temperature dependence of $\eta/s$. 

%
%

A supplemental approach for the extraction of $\eta/s$ is to leverage the longitudinal broadening of the 
transverse momentum two-particle correlation function $C_{2}$ \cite{Gavin:2006xd,Sharma:2008qr}:
\begin{eqnarray} \label{eq:C2}
C_{2}\left(\Delta\eta,\Delta\varphi \right) &=& \frac{ \left\langle \sum\limits_{\text{i}}^{n_1} \sum\limits_{\text{j} \neq \text{i}}^{n_{2}}
p_{\text{T},\text{i}} \; p_{\text{T},\text{j}}\right\rangle_{\eta_1,\varphi_1,\eta_2,\varphi_2}}{\langle n_{1}\rangle_{\eta_1,\varphi_1} \langle n_{2}\rangle_{\eta_2,\varphi_2}} \nonumber\\
&-& \langle p_{\rm T,1}\rangle_{\eta_1,\varphi_1} \langle p_{\rm T,2}\rangle_{\eta_2,\varphi_2},
\end{eqnarray}
where $\Delta\eta=\eta_{1}-\eta_{2}$ is the pseudorapidity difference between particles in bins $\eta_1$ and $\eta_2$ and $\Delta\varphi = \varphi_{1}-\varphi_{2}$ is the azimuthal angle difference between particles in bins $\varphi_{1}$ and $\varphi_{2}$; 
$n_1\equiv n(\eta_1,\varphi_1)$ and $n_2\equiv n(\eta_2,\varphi_2)$ are the event-wise number of charged particles in bins $(\eta_1,\varphi_1)$ and $(\eta_2,\varphi_2)$ and $p_{T,i}$ and $p_{T,j}$ are the transverse momenta of the i$^{th}$ and j$^{th}$ particles in their respective bins; the brackets $\langle ~\rangle$ represents event averages.

The $C_{2}\left(\Delta\eta,\Delta\varphi\right)$ correlator  reflects the covariance of the momentum currents \cite{Gavin:2006xd}. Therefore, it is sensitive to the dissipative viscous effects that ensue during the transverse and longitudinal expansion of the collisions' medium. Because such dissipative effects are more prominent for long-lived systems, they lead to longitudinal (or pseudorapidity) broadening of  $C_{2}(\Delta\eta, \Delta\varphi)$ as collisions become more central \cite{Sharma:2008qr}. 
 A proposed estimate of this broadening, $\Delta\sigma^2$,  can be linked to $\eta/s$ as \cite{Gavin:2006xd,Sharma:2008qr};
\begin{eqnarray} \label{eq:sigmacentral}
  \Delta \sigma^{2} &=& \sigma_{\rm c}^{2} - \sigma_{0}^{2} \nonumber \\
                              &=& \frac{4}{T_{\rm c}} \,\frac{\eta}{s}\,\left(\frac{1}{\tau_{0}} - \frac{1}{\tau_{\rm c,f}}\right),
\end{eqnarray}
where $\sigma_{\rm c}$ is the longitudinal width of $C_{2}\left(\Delta\eta \right)$ in central collisions and $\sigma_{0}$  is the longitudinal width at the formation time $\tau_{0}$; $T_{\rm c}$ and $\tau_{\rm c,f}$ represent the critical temperature  and the freeze-out time in central collisions. The freeze-out time  can be  constrained via the longitudinal femtoscopic radius, $R_{\rm long}$ \cite{Adamczyk:2014mxp,Adare:2014qvs,Aamodt:2011mr}.

In this work we investigate the utility of Eq.~\ref{eq:sigmacentral} for extraction of $\eta/s$ by using the events generated for Au+Au  collisions at $\sqrt{s_{NN}}$= 200 GeV, with the Heavy-Ion Jet Interaction Generator (HIJING) \cite{Wang:1991hta} and the AMPT model \cite{Lin:2004en}, to extract and study the longitudinal broadening of $C_{2}\left(\Delta\eta,\Delta\varphi\right)$ as a function of collision centrality for different input values of $\eta/s$. Here, it is noteworthy that the reliable extraction of $C_{2}\left(\Delta\eta,\Delta\varphi\right)$ requires robust suppression of possible self-correlations introduced via the use of the Particles Of Interest (POI) to define the collision centrality.

The current study is performed with simulated events for Au+Au collisions at $\sqrt{s_{NN}}$ = 200~GeV, obtained with the HIJING and the AMPT (v2.26t9b)~\cite{Lin:2004en} models with approximately 7~M and 5~M events respectively. In both models, charged particles with $0.2 < p_T < 2.0$ GeV/$c$, and $|\eta| < 1.0$ were selected for analysis. The HIJING model is employed to emphasize the effects of mini-jets, while the 
AMPT model allows for seamless variation of the input magnitude of $\eta/s$. 
The AMPT model, which has been widely employed to study relativistic heavy-ion  collisions \cite{Lin:2004en,Ma:2016fve,Ma:2013gga,Ma:2013uqa,Bzdak:2014dia,Nie:2018xog,Haque:2019vgi,Zhao:2019kyk,Bhaduri:2010wi,Nasim:2010hw,Xu:2010du,Magdy:2020bhd,Guo:2019joy,Magdy:2020gxf}, includes several important model ingredients: (i) an initial partonic state produced by the HIJING model~\cite{Wang:1991hta,Gyulassy:1994ew}, 
the values 
$a=0.55$ and $b=0.15$ GeV$^{-2}$ are used in the HIJING model for the Lund string fragmentation function 
$f(z) \propto z^{-1} (1-z)^a
\exp (-b~m_{\perp}^2/z)$, 
where $z$ represents the light-cone momentum 
fraction of the generated hadron of transverse mass $m_\perp$ with
respect to that of the fragmenting string. Also, (ii) partonic scattering with cross section,
\begin{eqnarray} \label{eq:21}
\sigma_{pp} &=& \dfrac{9 \pi \alpha^{2}_{s}}{2 \mu^{2}},
\end{eqnarray}
where $\alpha_{s}$ is the QCD coupling constant and $\mu$ is the screening mass. This cross section 
drives the expansion dynamics~\cite{Zhang:1997ej}; (iii) handronization via coalescence  followed 
by hadronic interactions~\cite{Li:1995pra}.  
 For a quark-gluon plasma of massless quarks and gluons
at a given temperature $T$, the input value of $\eta/s$   can be varied via an appropriate choice
of $\mu$ and/or $\alpha_s$ \cite{Xu:2011fi};
\begin{eqnarray} \label{eq:22}
 \dfrac{\eta}{s} &=& \dfrac{3 \pi}{40 \alpha^{2}_{s}}  \dfrac{1}{ \left(  9 +  \dfrac{\mu^2}{T^2} \right)  \ln\left(\dfrac{18 + \mu^2/T^2}{ \mu^2/T^2 } \right) - 18},
\end{eqnarray}
 For the present study we fix $\alpha_{s}$ = 0.47 and vary $\eta/s$ over the range 0.1--0.3 via an appropriate variation of $\mu$ for  $T$ = 378 MeV~\cite{Xu:2011fi}.


The events produced by the  HIJING and the AMPT models were analyzed with the longitudinal two-particle transverse momentum correlation function $C_{2}$ \cite{Gavin:2006xd,Sharma:2008qr};
\begin{eqnarray} \label{eq:23}
C_{2}\left(\eta_1,\varphi_1,\eta_2,\varphi_2 \right)  &=& 
\frac{\left\langle \sum\limits_{\text{i}}^{n_1} \sum\limits_{\text{j} \neq \text{i}}^{n_{2}}  p_{\text{T},\text{i}} \; p_{\text{T},\text{j}}\right\rangle}{\langle n_{1} \rangle \langle n_{2} \rangle} \nonumber \\
&-& \frac{\left\langle \sum\limits_{\text{i}}^{n_1} p_{\text{T},\text{i}} \right\rangle  \left\langle \sum\limits_{\text{i}}^{n_2} p_{\text{T},\text{i}} \right\rangle}{\langle n_{1} \rangle \langle n_{2} \rangle} 
\end{eqnarray}
Following Eq. \ref{eq:C2},  one can rewrite Eq.~\ref{eq:23} as,
\begin{eqnarray} \label{eq:24}
C_{2}\left(\Delta\eta,\Delta\varphi \right) &=& \frac{ \left\langle \sum\limits_{\text{i}}^{n_1} \sum\limits_{\text{j} \neq \text{i}}^{n_{2}}
p_{\text{T},\text{i}} \; p_{\text{T},\text{j}}\right\rangle_{\eta_1,\varphi_1,\eta_2,\varphi_2}}{\langle n_{1}\rangle_{\eta_1,\varphi_1} \langle n_{2}\rangle_{\eta_2,\varphi_2}} \nonumber\\
&-& \langle p_{\rm T,1}\rangle_{\eta_1,\varphi_1} \langle p_{\rm T,2}\rangle_{\eta_2,\varphi_2}.
\end{eqnarray}

The first term of Eq.~\ref{eq:23}  can be rewritten as,
\begin{eqnarray} \label{eq:25}
 \frac{  \left\langle \sum\limits_{\text{i}}^{n_1} \sum\limits_{\text{j} \neq \text{i}}^{n_{2}} 
        p_{\text{T},\text{i}} \; p_{\text{T},\text{j}}\right\rangle  }{\langle n_{1} \rangle \langle n_{2} \rangle}
        &=&
         \frac{  \left\langle \sum\limits_{\text{i}}^{n_1} \sum\limits_{\text{j} \neq \text{i}}^{n_{2}} 
        p_{\text{T},\text{i}} \; p_{\text{T},\text{j}}\right\rangle  }{\left\langle \sum\limits_{\text{i}}^{n_1} \sum\limits_{\text{j} \neq \text{i}}^{n_{2}} 
        n_{\text{i}} \; n_{\text{j}}\right\rangle} r_{1,2},
\end{eqnarray}
where,
\begin{eqnarray} \label{eq:26}
r_{1,2} &=& \dfrac{\left\langle \sum\limits_{\text{i}}^{n_1} \sum\limits_{\text{j} \neq \text{i}}^{n_{2}} 
        n_{\text{i}} \; n_{\text{j}}\right\rangle }{\langle n_{1} \rangle \langle n_{2} \rangle}.
\end{eqnarray}
Note that $r_{1,2}$ is a number correlation that will be unity when the particle pairs are independent (i.e., the pair distribution will factor into the product of single-particle distributions). Experimentally the $r_{1,2}$ correlations can be impacted by (i) the detector acceptance and (ii) the centrality definition. The latter is especially important for measurements which employ charged particle multiplicity to calibrate the collision centrality. That is, if the particles used to determine $C_{2}\left(\Delta\eta,\Delta\varphi \right)$
are also used to define the  event centrality, a residual self-correlation appears in $r_{1,2}$ which can serve to distort the correlator and hence, the value of the extracted longitudinal broadening. 
Such an effect can be reduced by separating the particles used to to determine $C_{2}\left(\Delta\eta,\Delta\varphi \right)$ 
and those use in the centrality definition \cite{Acharya:2019oxz}.
{\color{black}
This can also be achieved via randomly sampling 50\% of the charged particles in an event to be used for the centrality definition and the other 50\% to participate in measuring the $C_{2}\left(\Delta\eta,\Delta\varphi \right)$. The latter method could be used experimentally when the detector has a limited $\eta$ acceptance. The efficacy of this method can be studied in models by comparing the results with those using the impact parameter to define centrality. 
}
\begin{figure*}[hbt]
    \includegraphics[width=1.0 \linewidth, angle=-0,keepaspectratio=true,clip=true]{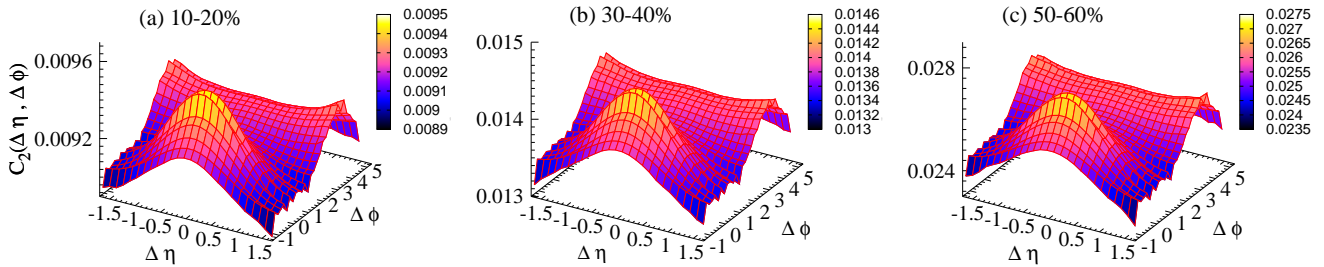}
     \vskip -0.2cm
    \caption{
    Comparison of the $C_{2}\left(\Delta\eta,\Delta\varphi \right)$ correlators obtained from 10-20\%, 30-40\% and 50-60\% central HIJING events for Au+Au collisions at 200 GeV. }\label{fig:1}
		 \vskip -0.2cm
\end{figure*}

\begin{figure*}[ht] 
    \includegraphics[width=1.0 \linewidth, angle=-0,keepaspectratio=true,clip=true]{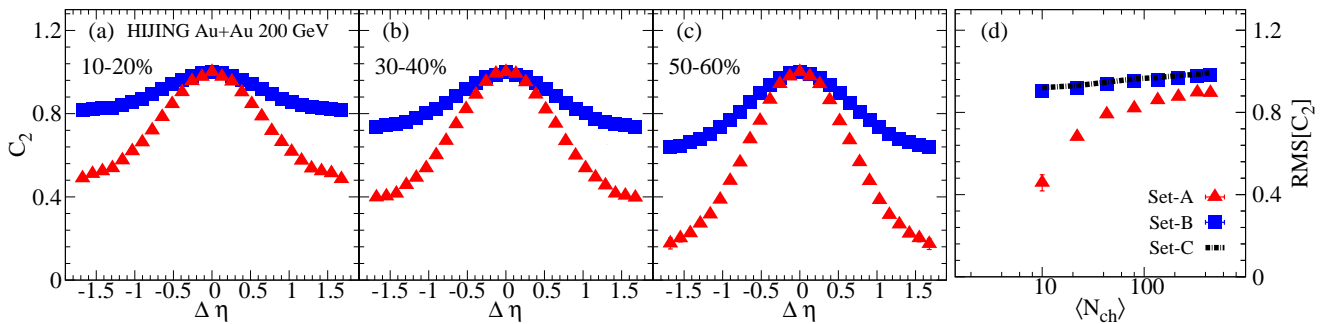}
     \vskip -0.2cm
    \caption{
 Comparison of the simulated $C_2(\Delta\eta)$ correlators for 10-20\% (a), 30-40\% (b) and 50-60\% (c)
Au+Au collisions at 200 GeV, obtained with the HIJING model. The Set-$A$ distributions are for 
centrality defined using all charged particles in an event; The Set-$B$ distributions are for
centrality defined using a random sampling of charged particles in an event. The distributions are peak normalized.
Panel (d) compares the centrality-dependent RMS values for set-$A$, set-$B$ and set-$C$ (centrality defined using the impact parameter distribution).
  } \label{fig:2}
  \vskip -0.2cm
\end{figure*}
\begin{figure}[hbt]
\includegraphics[width=0.85 \linewidth, angle=-0,keepaspectratio=true,clip=true]{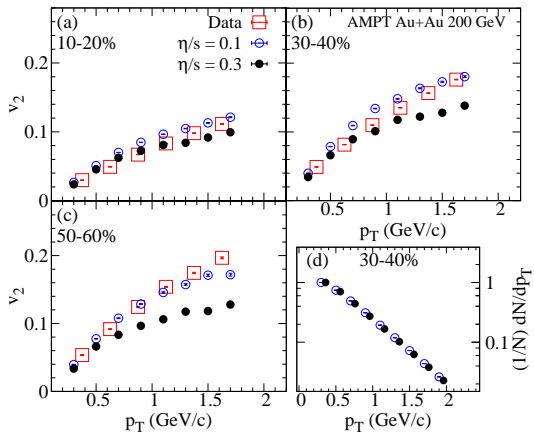}
 \vskip -0.3cm
 \caption{
Comparison of the experimental and simulated elliptic flow $v_2(p_T)$, for 10-20\% (a), 30-40\% (b), and 50-60\% (c) central Au+Au collision at 200 GeV. The open markers indicate the experimental data~\citep{Adare:2011tg} and the solid markers represent the results from AMPT model calculations with different values of $\eta/s$. In panel (d) we show the $p_{T}$ distribution for the 30-40\% central Au+Au collision at 200 GeV, {\color{black}they indicated about 9\% difference between the two sets.} 
} \label{fig:3} 
 \vskip -0.3cm
\end{figure}
\begin{figure*}[hbt]
\includegraphics[width=0.9 \linewidth, angle=-0,keepaspectratio=true,clip=true]{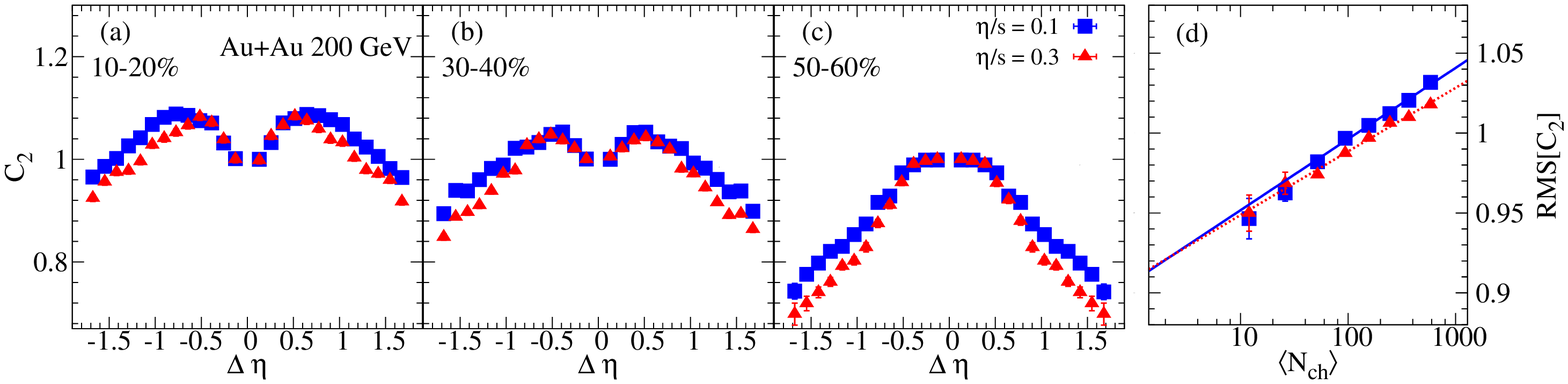} 
 \vskip -0.3cm
     \caption{
		Comparison of the simulated $C_2(\Delta\eta)$ correlators for 10-20\% (a), 30-40\% (b) and 50-60\% (c)
Au+Au collisions at 200 GeV, obtained with the AMPT model {\color{black}using set-B of centrality definition.} Results are compared for two values of $\eta/s$ as indicated; the distributions are peak-normalized. Panel (d) compares the extracted centrality-dependent RMS values for the two indicated values of $\eta/s$.
}\label{fig:4}
 \vskip -0.3cm
\end{figure*}
%

%
The influence of the self-correlations were studied using HIJING events. First, $C_{2}\left(\Delta\eta,\Delta\varphi \right)$ was generated for these events;
Fig.~\ref{fig:1}  shows representative correlators  for 10-20\%, 30-40\%,  and 50-60\% central 
Au+Au collisions. They indicate a sizable near side peak which show a weak centrality dependence.
Second, in Fig.~\ref{fig:2} the $C_{2}\left(\Delta\eta,\Delta\varphi \right)$ correlators were projected onto the $\Delta\eta$ axis for $| \Delta\phi | < 1.0$, to facilitate the extraction of the RMS of $C_{2}\left(\Delta\eta\right)$ \cite{Agakishiev:2011fs}.	
Subsequently, the set of RMS values [for $C_2(\Delta\eta)$] for the respective centrality selections 
were obtained for three separate centrality definitions;
(i) Set-A uses charged particles within our kinematic cuts for both centrality determination and C2 calculations, (ii) Set-B uses random sampling of the charged particles in an event to  suppress self correlations, and (iii) Set-C uses centrality defined using the impact parameter distribution to suppress self correlations.

Figures \ref{fig:2} (a)-(c) compare the centrality-dependent $C_2(\Delta\eta)$  distributions for Set-A and Set-B.  The comparison indicates significant differences in the RMS values for the respective distributions. This difference is more transparent in Fig.~\ref{fig:2} (d)  which compares the extracted RMS values for Set-A, Set-B and Set-C.  The RMS values for Set-B and Set-C show the expected agreement consistent with the absence of significant self-correlations.
They also confirm that excluding the POI from the collision centrality definition, serves to reduce possible self-correlations. 
 Note as well the characteristic dependence of the RMS values on the charged particle multiplicity $N_{\rm ch}$, i.e., the RMS values show a linear increase with $\log(\langle N_{ch} \rangle)$, as observed experimentally \citep{Acharya:2019oxz}. 
Set-A shows significant deviations from the true RMS values due to self-correlations introduced by using the POI in the definition of the collision centrality.  Therefore, the suppression or avoidance of such self-correlations in experimental measurements, is crucial for reliable measurements of the longitudinal broadening of $C_2(\Delta\eta)$.

The AMPT model was employed in our study of the influence of $\eta/s$ on the longitudinal broadening 
of $C_2(\Delta\eta)$ \cite{Acharya:2019oxz}. For these simulations, $\mu$ was varied [with $\alpha_{s}$ = 0.47 and $T$ = 378 MeV~\cite{Xu:2011fi}] in conjunction with Eq.~\ref{eq:22} to obtain simulated results for $\eta/s$ =0.1, 0.2 and 0.3.
Charged particles with $0.2 < p_T < 2.0$ GeV/$c$, and $|\eta| < 1.0$ were used in the analysis and 
the POI were excluded from the collision centrality definition. 

The influence of $\eta/s$ was benchmarked by first comparing the experimental $v_2(p_T)$ values \citep{Adare:2011tg} with the simulated ones obtained with AMPT for $\eta/s$= 0.1 and 0.3~\cite{Nasim:2016rfv}.
Fig.~\ref{fig:3} illustrates such a comparison for 10-20\%, 30-40\%,  and 50-60\%  Au+Au collisions. 
Here, the essential point is that the simulated values of $v_2(p_T)$ show a clear sensitivity to the 
magnitude of $\eta/s$, as well as the expected decrease in the magnitude of $v_2(p_T)$ when $\eta/s$ is increased. 
Here one can think that varying the model $\eta/s$ could impact the transverse radial velocity profile.  Such an effect can be addressed by comparing the $p_{T}$ distribution for different values of $\eta/s$. In panel (d) we are showing the $p_{T}$ distribution for the 30-40\% central Au+Au collision at 200 GeV for $\eta/s$ = 0.1 and 0.3. The $p_{T}$ distributions show a small change ($<~9\%$) when $\eta/s$ vary from 0.1 to 0.3.

\begin{figure}[hbt]
    \centering
\includegraphics[width=0.60 \linewidth, angle=-90,keepaspectratio=true,clip=true]{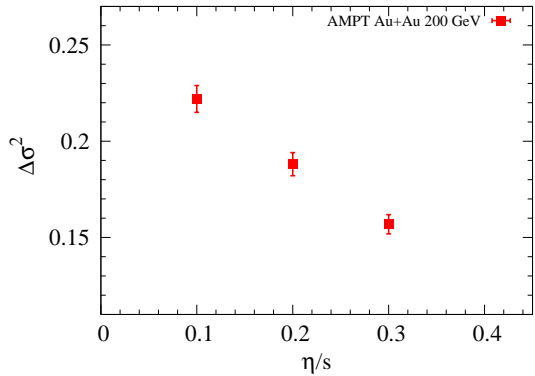}
 \vskip -0.2cm
     \caption{
      $\Delta\sigma^{2} = \sigma_{\rm c}^{2} - \sigma_{0}^{2} $  vs. $\eta/s$  for 200 GeV Au+Au collisions simulated with AMPT model.     
      }
    \label{fig:5} 
 \vskip -0.3cm
\end{figure}

Following the benchmarking of $\eta/s$, the distributions for $C_2(\Delta\eta)$ were extracted and used to determine the centrality-dependent RMS values. Figs. \ref{fig:4} (a) - (c) compare representative $C_2(\Delta\eta)$ distributions for   $\eta/s$ = 0.1 and 0.3. They indicate narrower distributions for the correlators obtained with $\eta/s$ = 0.3.
The RMS values extracted from such distributions for a more detailed set of centrality selections, are shown as a function of $N_{\rm ch}$ in Fig. \ref{fig:4} (d). They indicate the characteristic increase of the RMS values with $\log(\langle N_{\rm ch}\rangle)$ but with a smaller slope for $\eta/s = 0.3$. 
This reduction in the magnitude of the RMS values with increasing $\eta/s$ suggests that, in the AMPT model, an increase in $\eta/s$ (for a fixed value of $\alpha_{s}$) serves to dampen the magnitude of the longitudinal fluctuations.

To further investigate the influence of $\eta/s$ on the longitudinal broadening, we estimated 
$\Delta\sigma^2$ from the simulated centrality-dependent data set for each value of $\eta/s$ as follows.
First, $\sigma^2$ was extracted and plotted as a function of centrality for each value of $\eta/s$. Second,  for a given value of $\eta/s$
the difference of the RMS values ($\Delta\sigma^2$) for the most central and the most peripheral collisions was obtained. Fig.~\ref{fig:5} shows the dependence of $\Delta\sigma^{2}$ on $\eta/s$. It indicates a monotonic decrease of $\Delta\sigma^{2}$ with $\eta/s$ over the range studied. 
This observed sensitivity of  $\Delta\sigma^{2}$ to the magnitude of $\eta/s$, suggests that 
experimental studies of the longitudinal broadening of the transverse momentum two-particle correlator could provide additional constraints for precision extraction of $\eta/s$.

  
In summary, we have studied the sensitivity of the longitudinal broadening of the transverse momentum two-particle correlator $C_{2}(\Delta\eta)$, to varying degrees of specific viscosity using the AMPT model.  We find that reliable extraction of $C_{2}(\Delta\eta)$ requires the avoidance or suppression of self-correlations associated with the common practice of including the particles used to determine $C_{2}(\Delta\eta)$ in the set used to characterize the centrality.
The RMS values extracted from the $C_{2}(\Delta\eta)$ distributions,  indicate a characteristic linear dependence on $\log(\langle N_{ch}\rangle)$ with slopes that vary with $\eta/s$. Furthermore, the estimated values of the longitudinal broadening $\Delta\sigma^{2}$,  indicate a monotonic decrease with $\eta/s$. This sensitivity to $\eta/s$, suggests that new experimental measurements of the longitudinal broadening of the transverse momentum two-particle correlator could provide additional constraints for precision extraction of $\eta/s(T)$.
%
  
\section*{Acknowledgments}
The authors thank Emily E. Racow for the useful discussion.
This research is supported by the US Department of Energy under contract DE-FG02-94ER40865 (NM) and DE-FG02-87ER40331.A008 (RL).

\bibliography{ref} 
\end{document}